\documentclass[pdflatex,sn-mathphys-num]{sn-jnl}

\usepackage{graphicx}%
\usepackage{multirow}%
\usepackage{amsmath,amssymb,amsfonts}%
\usepackage{amsthm}%
\usepackage{mathrsfs}%
\usepackage[title]{appendix}%
\usepackage{xcolor}%
\usepackage{pbox}
\usepackage{textcomp}%
\usepackage{manyfoot}%
\usepackage{algorithm}%
\usepackage{algorithmicx}%
\usepackage{algpseudocode}%
\usepackage{listings}%
\usepackage{tikz}
\usepackage{caption}
\usepackage{float}
\usepackage{booktabs}
\usepackage{array}
\usepackage{siunitx}
\usepackage{tikz-feynman}
\usetikzlibrary{positioning,arrows}
\usetikzlibrary{decorations.pathmorphing}
\usetikzlibrary{decorations.markings}
\usetikzlibrary{shapes.geometric}
\usepackage{endnotes}
\tikzset{
    vector/.style={decorate, decoration={snake}, draw},
    provector/.style={decorate, decoration={snake,amplitude=2.5pt}, draw},
    antivector/.style={decorate, decoration={snake,amplitude=-2.5pt}, draw},
    fermion/.style={draw=black,
      postaction={decorate},decoration={markings,mark=at position .55
        with {\arrow[draw=black]{>}}}},
    fermionbar/.style={draw=black, postaction={decorate},
                       decoration={markings,mark=at position .55 with {\arrow[draw=black]{<}}}},
    fermionnoarrow/.style={draw=black},
    gluon/.style={decorate, draw=black,decoration={coil,amplitude=4pt, segment length=6pt}},
    scalar/.style={dashed,draw=black,
      postaction={decorate},decoration={markings,mark=at position .55
        with {\arrow[draw=black]{>}}}},
    scalarbar/.style={dashed,draw=black,
      postaction={decorate},decoration={markings,mark=at position .55
        with {\arrow[draw=black]{<}}}},
    scalarnoarrow/.style={dashed,draw=black},
    electron/.style={draw=black,
      postaction={decorate},decoration={markings,mark=at position .55
        with {\arrow[draw=black]{>}}}},
    bigvector/.style={decorate, decoration={snake,amplitude=4pt}, draw},
}

\theoremstyle{thmstyleone}%
\theoremstyle{thmstyletwo}%

\theoremstyle{thmstylethree}%
\begin{document}

\title[Article Title]{Estimating power corrections for the Drell-Yan Process}

\author[1]{\fnm{Ekta} \sur{Chaubey}}\email{eekta@uni-bonn.de}
\equalcont{These authors contributed equally to this work.}

\author*[2]{\fnm{Pooja} \sur{Mukherjee}}\email{pooja.mukherjee@desy.de}

\affil[1]{Bethe Center for Theoretical Physics, Universität Bonn, 53115 Bonn, Germany}
\affil*[2]{II. Institut für Theoretische Physik, Universität Hamburg, Luruper Chaussee 149, 22761, Hamburg,
Germany}


\abstract{We study power corrections in the Drell-Yan (DY) process using state-of-the-art predictions for both neutral and charged current production. For both types of DY processes, we account for power corrections arising from bottom and charm quark effects within a variable flavor number scheme. Our results show that these corrections become significant in the low-$Q$ region. We also ensure proper treatment of overlapping contributions by carefully applying matching procedures to eliminate any double counting.}




\maketitle

\section{Introduction}\label{sec1}

High-energy hadron-hadron colliders offer unique opportunities to probe fundamental physics, from searches for beyond-the-Standard Model particles to precision studies of known Standard Model (SM) processes. A key benchmark process in this context is the Drell-Yan (DY) process, which facilitates lepton pair production via neutral-current (NC) and charged-current (CC) interactions. DY measurements across a wide range of collider energies have provided critical insights into hadron structure, SM parameters, and the validity of factorization theorems. The LHC, with its unprecedented precision, has further refined our understanding, and its upcoming high-luminosity phase promises even greater improvements. These experimental advancements, in turn, drive the need for increasingly precise theoretical predictions. This need translates to a complete understanding of the DY process for future precision era. For a comprehensive review of the state-of-the-art QCD corrections to this process, see~\cite{Chaubey:2025lan}. Of course, QED corrections also play a crucial role, as they significantly affect the determination of the lepton-pair transverse momentum spectrum and compete with quark mass effects; see, for example,~\cite{Cieri:2018sfk,Autieri:2023xme}. A related phenomenological analysis focusing on an improved description of the bottom quark is presented in~\cite{Bagnaschi:2018dnh}.

Recent progress includes the calculation of the inclusive DY cross-section up to $\text{N}^3\text{LO}$ in perturbative QCD, as well as differential predictions incorporating electroweak(EW) and mixed QCD-EW corrections, quark mass effects, and low-transverse momentum dynamics. Parton distribution function (PDF) extractions have also reached new levels of accuracy, with partial $\text{N}^3\text{LO}$ results available. A particularly interesting, but unexplored kinematic regime at the LHC is the low invariant mass and forward rapidity region, accessible via LHCb. This region is sensitive to small $x$ (parton momenta fractions) of PDFs at low virtuality but also presents theoretical challenges due to slower perturbative convergence of partonic cross sections and potential quark mass effects. 

In this contribution, we estimate the power corrections to the inclusive NCDY and CCDY cross sections using state-of-the-art theoretical predictions. We also outline the method used to extract these power corrections within the massive variable flavour number scheme (MVFNS). This contribution serves as a prelude to  \cite{Chaubey:2025lan}, where a comprehensive analysis is presented. In particular, the sources of uncertainty in DY are examined in detail for both NC and CC processes, with particular emphasis on the low invariant mass region. Differential predictions for the invariant mass spectrum are provided at N$^3$LO, supplemented by exact charm and bottom quark mass effects at $\mathcal{O}(\alpha_s^2)$. Further investigations on the impact of PDF choices (including approximate N$^3$LO sets), scale variations, the strong coupling constant, and heavy-quark mass effects on the resulting distributions is also carried out.

The writeup is structured as follows: Section 2 introduces the computational frame-work for both NC and CC DY processes, detailing the theoretical fixed-order calculations and the matching procedure employed
to combine massive and massless calculations. Section 3 presents the results of the MVFNS corrections, before the conclusions in section 4.

\section{Theoretical Setup}
We start with theoretical set up of  DY process. Let there be two protons (P) scattering with momenta $P_1$ and $P_2$, and let $Q$ be the invariant mass of the final state particles $L_1 L_2$. Then the process is given as:
\begin{align}
    {\rm{P}}(P_1) + {\rm{P}} (P_2) \rightarrow L_1 L_2 (Q)+ X.
\end{align}
where $L_1L_2$ are lepton pairs ($l^{+}l^{-}$) for NCDY and lepton-neutrino pair ($l^{\pm}\nu_l$) for CDY.
Here, $X$ denotes the additional QCD radiation.  The cross section is mediated through a virtual $Z$ boson or a photon in case of NCDY and a virtual $W$-boson for CCDY to decay subsequently into a  pair of lepton (lepton-neutrino for CCDY) for NCDY in the final state.

Let $S= {(P_1+P_2)}^2$ be the incoming energy of the protons. The hadronic differential cross section for DY production with respect to $Q$ can be written using the QCD factorization theorem as 
\begin{equation}
Q^2 \; \frac{d \sigma^{(n_f, \kappa)}}{dQ^2}= \tau \sum_{i,j} \mathcal{L}_{ij}^{(n_f)} (\tau, \mu_F) \otimes \eta _{ij}^{(n_f,\kappa)}(\tau, a_s^{(n_f)}(\mu_R)).
\label{eq:4fsdiffcrosssection}
\end{equation}
Here $\tau$ is the hadronic scaling variable, defined as 
    $\tau= \frac{Q^2}{S}$ and $\kappa\in\{\pm1,0\}$, depending on whether the process is mediated by a neutral vector boson ($\kappa=0$) or a $W^{\pm}$ ($\kappa=\pm1$). $\mu_F$ and $\mu_R$ denote the factorization scale and the renormalization scale, respectively. The $\eta_{ij}^{(n_f,\kappa)}$ denotes the partonic cross section for producing a lepton pair or a lepton-neutrino pair from a collision of two partons $i$ and $j$. The sum runs over all active partons in the proton, i.e., gluons and all massless quark flavors. The partonic luminosity $\mathcal{L}_{ij}^{(n_f)}$ is the convolution over the $n_f$-flavor PDFs:
\begin{align}
\mathcal{L}_{ij}^{(n_f)} (\tau, \mu_F)&= f_i^{(n_f)} (\tau, \mu_F) \otimes f_j^{(n_f)} (\tau,\mu_F) \nonumber \\ &\equiv \int_0^1 f_i^{(n_f)} (x_1, \mu_F)\; f_j^{(n_f)} (x_2, \mu_F)\; \delta(\tau- x_1 x_2)\; dx_1 dx_2.
\label{eq: etafifj}
\end{align}
The $f_i^{(n_f)}(x_i, \mu_F)$ represents the PDFs, i.e., the probability of finding a parton $i$ with momentum fraction $x_i$ inside the proton. The partonic cross section $\eta_{ij}^{(n_f, \kappa)}$ exhibits a perturbative expansion in the strong coupling $a_s$, with $a_s= \frac{\alpha_s}{\pi}$, given by
\begin{align}
    \eta _{ij}^{(n_f,\kappa)} (\tau, a_s^{(n_f)}(\mu_R)) = \sum_{k = 0}^\infty a_s^{(n_f) k} (\mu_R) \;\eta _{ij}^{(n_f, \kappa, k)} (\tau).
    \label{eq:eta4}
\end{align}

\subsection{The Matching Procedure for Drell-Yan Process}
\label{sec:matching_theory}
In this section, we discuss the matching procedure for DY process. We start by discussing the NCDY process. In matrix element computations involving bottom quarks, they can be treated as either massless or massive. The 5-flavour scheme (5FS) assumes all quarks are massless, including the bottom quark, which has an associated PDF and contributes to QCD evolution with 
$n_f=5$. However, neglecting the bottom mass in final-state processes involving massive particles like the Z boson introduces infrared and collinear singularities, as mass effects are resummed inside the PDFs in the 5FS cross section.

Alternatively, the 4-flavour scheme (4FS) treats the bottom quark as massive and the 3-flavour scheme (3FS) treats both bottom and charm quarks massive. This means that they appear only in the final states and have no PDFs. In these schemes, both bottom and charm quarks arise from gluon splittings at the leading order, and all finite-mass effects are included in the partonic cross section. While this avoids collinear divergences, it introduces large logarithms of the form 
$\log \frac{Q^2}{m_Q^2}$  which can spoil perturbative convergence, where $m_Q \in \{m_c,m_b\}$. These logarithms are resummed in the 5FS, improving predictive accuracy. Representative diagrams contributing to various orders in perturbation theory for 4FS is shown in Tables~\ref{fig:diagrams}.
\begin{table}[!htbp]
\begin{center}
\resizebox{13cm}{!}{
\begin{tabular}{|c|c|c|c|}
\hline\hline
 & \begin{tikzpicture}[line width=1 pt, scale=0.7]
\draw[fermion] (-2.5,1) -- (0,0);
\draw[fermion](0,0)  -- (-2.5,-1);
\draw[vector] (0,0) -- (2,0);
\node at (-2.7,1) {$b$};
\node at (-2.7,-1) {$\bar{b}$};
\node at (2.5,0.5) {\scriptsize{$\gamma^{*}/Z^{*}\rightarrow l^{+}l^{-}$}};
\end{tikzpicture} & 
\begin{tikzpicture}[line width=1 pt, scale=0.7]
\draw[gluon] (-2.5,1) -- (0,1);
\draw[fermion] (-2.5,-1) -- (0,-1);
\draw[fermion] (0,-1) -- (0,1);
\draw[fermion] (0,1) -- (2,1);
\draw[vector] (2,-1) -- (0,-1);
\node at (-2.7,1) {$g$};
\node at (-2.7,-1) {$b$};
\node at (2.2,1) {$b$};
\node at (2.2,-0.5) {$\gamma/Z\rightarrow l^{+}l^{-}$};
\end{tikzpicture} & 
\begin{tikzpicture}[line width=0.5 pt, scale=0.7]
\draw[gluon] (-2.5,1) -- (0,1);
\draw[gluon] (-2.5,-1) -- (0,-1);
\draw[fermion] (0,-1) -- (0,1);
\draw[vector] (0,0) -- (2,0);
\draw[fermion] (0,1) -- (2,1);
\draw[fermion] (2,-1) -- (0,-1);
\node at (-2.7,1) {$g$};
\node at (-2.7,-1) {$g$};
\node at (2.2,1) {$b$};
\node at (2.2,-1) {$\bar{b}$};
\node at (2.5,0.5) {$\gamma/Z\rightarrow l^{+}l^{-}$};
\end{tikzpicture}
 \begin{tikzpicture}[line width=0.5 pt, scale=0.7]
\draw[fermion] (-2.5,1) -- (0,1);
\draw[fermion] (0,-1) --(-2.5,-1) ;
\draw[fermion] (0,1) -- (0,-1);
\draw[gluon] (0,1) -- (2,1);
\draw[fermion] (2,1) -- (3,1.5);
\draw[fermion] (3,0.5) -- (2,1);
\draw[vector] (2,-1) -- (0,-1);
\node at (-2.7,1) {$q$};
\node at (-2.7,-1) {$\bar{q}$};
\node at (3.2,1.5) {$b$};
\node at (3.2,0.5) {$\bar{b}$};
\node at (2.3,-0.5) {$\gamma/Z\rightarrow l^{+}l^{-}$};
\end{tikzpicture}
\\ \hline &&&\\
  $\quad$4FS$\quad$ & -- & -- & LO  \\
 $\quad$5FS$\quad$ & LO & NLO & NNLO \\
  \hline\hline
\end{tabular}}
\caption{\label{fig:diagrams}  Representative diagrams for the NCDY process at different orders in perturbation theory in the 4FS and 5FS.
}
 \end{center}
 \end{table}

The results in the massless limit are known up to $\text{N}^3\text{LO}$~\cite{Duhr:2021vwj}, whereas the massive ones have been computed up to NNLO in~\cite{Behring:2020uzq}. The matched calculation for Higgs production in bottom quark annihilation was performed using the FONLL scheme~\cite{Forte:2016sja} up to $\text{N}^3\text{LO}$ in~\cite{Duhr:2020kzd}. Studies on $Z$ boson production in bottom quark fusion using the same FONLL procedure were performed in~\cite{Forte:2018ovl}. The operator-matrix-elements (OMEs) used in this article were computed in~\cite{Buza:1996wv}. The matching procedure adopted in our article was first performed in~\cite{Gauld:2021zmq} for NCDY. 
\subsubsection{Estimating the Mass Corrections in DY}

In the MVFNS, we combine massless DY with the power correction terms extracted from the massive computations, as outlined below. For brevity, from now on we denote $a_s^{(n_f) k}$ as $a_s$. The cross-section in the MVFNS is defined as:
\begin{align}
    {\rm{d}} \sigma ^{\rm{MVFNS}} = {\rm{d}} \sigma ^{(5,\kappa)} + \sum_{i = c,b}^{n_f} {\rm{d}} \sigma_{i,pc}^{(5,\kappa)}.
    \label{eq:5MVFNS}
\end{align}
This construction of d$\sigma ^{\rm{MVFNS}}$ includes the resummation of massive collinear logarithms to all orders in $a_s$. Additionally, the exact heavy quark mass dependence is incorporated up to the fixed-order accuracy to which the power correction terms ${\rm{d}}\sigma_{i,pc}^{(5,\kappa)}$ is computed. The first term on the right-hand side, ${\rm{d}} \sigma ^{(5,\kappa)} $, is obtained from massless computations using {\tt{n3loxs}}, while the second term is derived using the procedure detailed below.

To provide a uniform definition of NCDY scattering across all energy scales, we combine the 3FS, 4FS and 5FS schemes. As an example for combining 4FS and 5FS cross sections for NCDY, we extract the power-suppressed terms from equation~\ref{eq:5MVFNS} as follows. The massive differential cross-section is expressed as a sum of three components:
\begin{align}
    {\rm{d}}\sigma^{(4,0)} = {\rm{d}}\sigma_{n_f}^{(5,0)} + {\rm{d}}\sigma _{\ln[m]}^{(5,0)} + {\rm{d}}\sigma_{pc}^{(4,0)}.
    \label{eq:3componentdiffcross}
\end{align}
Here, ${\rm{d}}\sigma^{(4,0)}$ represents the fully massive contribution from a single heavy-flavor of mass $m_{Q}$, where $Q$ could denote either a charm or bottom quark. The components on the right-hand side are:
\begin{itemize}
    \item \textbf{$n_f$ corrections:} ${\rm{d}}\sigma_{n_f}^{(5,0)}$ are the $n_f$ contributions arising from a single flavor in the massless limit. In the massless computation, the heavy quark $Q$ contributes to the same subprocesses as in the massive computation, but with $m=0$. Thus, these corrections can be directly extracted from the massless partonic cross-section at any given order in $\alpha_s$. We obtained these corrections from the publicly available {\tt{n3loxs}}~\cite{Baglio:2022wzu}. These terms diverge in the limit $m \rightarrow 0$.
    \item \textbf{Logarithmic corrections:} ${\rm{d}}\sigma _{\ln[m]}^{(5,0)}$ are the contributions from massive collinear logarithms. These can be obtained using the decoupling relations for the PDFs and $a_s$. The logarithmic dependence of the massive corrections originates from collinear divergences. These corrections are constructed using only massless inputs by expressing both results in terms of a common set of PDFs and $a_s$ in the 5FS scheme. 
    \item \textbf{Power corrections:} ${\rm{d}}\sigma_{pc}^{(5,0)}$ are contributions that vanish smoothly in the massless limit. These arise exclusively in the massive computations.
\end{itemize}


%

We briefly explain the decoupling relations used to describe parameters such as $\alpha_s$ and PDFs in a theory with a massive quark (4FS) relative to an effective theory where the corresponding quark is treated as massless (5FS) at a fixed order. The strong coupling constant relation between the 4FS and 5FS schemes, matched at scale $\mu_R$, is as given in~\cite{Chetyrkin:1997sg}. The PDF relation is provided in terms of OMEs ($A_{ij}$), which describe transitions between partonic states $i \rightarrow j$:
\begin{align}
f_i^{(5)} = \sum_{j=-4}^4 K_{ij}\big(L_Q, a_s^{(4)}\big) \otimes f_j^{(4)}, \quad -5 \leq i \leq 5,
\end{align}
where $K_{ij}$ are kernels from~\cite{Buza:1996wv}. Substituting this relation into above removes the $b$-quark PDF  from the initial states and gives rise to contributions to ${\rm{d}}\sigma _{\ln[m]}^{(5,0)}$. The contribution up to $O(a_s^2)$ reads as:
\begin{align}
{\rm{d}}\sigma _{\ln[m]}^{(5,0)} = \tau \sum_{i,j=-4}^4 \mathcal{L}_{ij}^{(5)}(\tau, \mu_F) \otimes \mathbf{A}_{ij}(\tau, \mu_F, \mu_R, L_Q, a_s, m),
\end{align}
where $\mathbf{A}_{ij} = \eta_{ij}^{(5,0,k)} + \delta \eta_{ij}^{(5,0,k)}$.  The expansion in $a_s^k$ for $\delta \eta_{ij}^{(n_f,0,k)}$ is:
\begin{align}
\delta \eta_{gg}^{(5,0,2)} &= 2 A_{bg}^{(1)} \otimes A_{bg}^{(1)} \otimes \eta_{q\bar{q}}^{(5,0,0)} + 4 A_{bg}^{(1)} \otimes \eta_{qg}^{(5,0,1)}.
\label{eq:powerexpansionofAs}
\end{align}
Similar generalisations to 5FS vs 3FS matching terms for NCDY are as follows:

 \begin{align}
\delta \eta^{(5,0,2)} _{gg} =& \sum_{i\in \{c,b \}} 2 A_{ig }^{(1)} \otimes A_{ig}^{(1)} \otimes \eta_{q \bar{q}}^{(5,0,0)} + 4 A_{ig}^{(1)} \otimes \eta _{q g}^{(5,0,1)}.
 \label{eq:powerexpansionofAs5vs3}
\end{align}
For CCDY ($W^-$ production), these equations attain the form
\begin{align}
    &\delta\eta_{g\bar u}^{(5,-1, 1)} = A_{bg}^{(1)}\otimes \eta_{b\bar u}^{(5,-1,0)},\nonumber\\&
    \delta\eta_{gi}^{(5,-1,1)} = A_{cg}^{(1)}\otimes \eta_{\bar c i}^{(5,-1,0)} , \qquad i \in \{ d,s\}
\end{align}
Here the one-loop OME is given as,
\begin{align}
    A_{Qg}^{(1)} &= -2 \bigg( z^2 + (1-z)^2 \bigg)\ln{\frac{m_Q^2}{\mu^2}}, \qquad Q \in \{ c,b\}
    \label{eq:defAQg}
\end{align}
All these convolutions are computed analytically using \textsc{PolyLogTools}~\cite{Duhr:2019tlz}, expressed in terms hyper-logarithmic (HPL) functions which are later on computed numerically using the computer algebra system {\tt{GiNaC}}~\cite{Bauer:2000cp}. The massive corrections are computed using in-house code for NCDY and {\tt MCFM}-10.3~\cite{Campbell:2015qma,Campbell:2019dru} for CCDY. For further clarifications on estimating the mass corrections in DY we refer the interested readers to~\cite{Chaubey:2025lan}
.
\section{Results}\label{sec_results}
\begin{figure}
     \centering
    \includegraphics[width=0.8\linewidth]{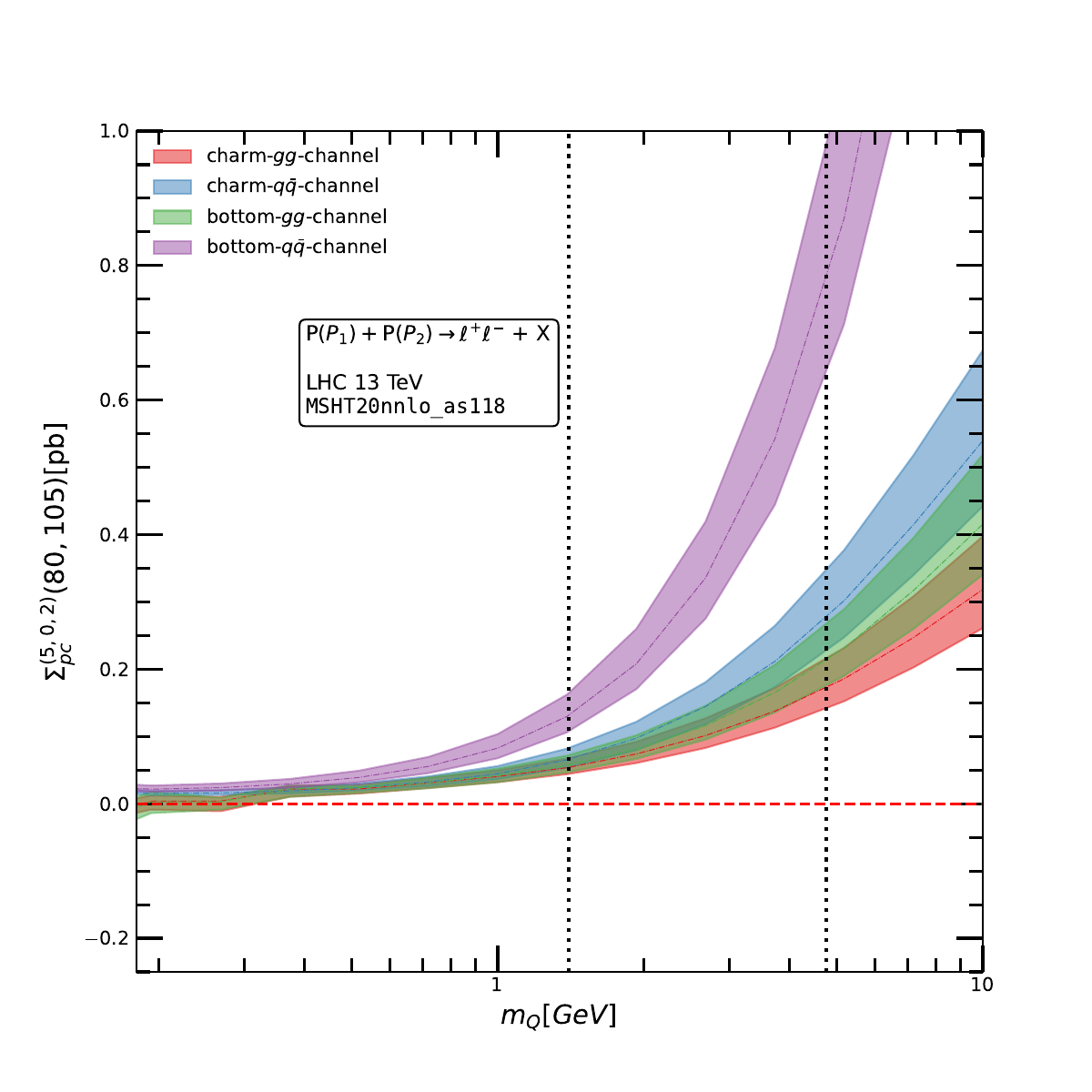}
    \caption{ Massive power-corrections to the NCDY cross-section in the $Q \in [80, 105]$ GeV. The bands represent the 7-point variation of $\mu_R$ and $\mu_F$ around the central (dynamic) scale $\mu_R=\mu_F=Q$. We observe that the power corrections vanish in the limit $m_Q \rightarrow 0$.}
     \label{fig:enter-label3}
 \end{figure}

In this section, we estimate the impact of power correction terms derived from 4FS and 3FS massive corrections. To present the phenomenological results for the binned invariant-mass distribution we use the following notation. Each term in eq.~\eqref{eq:3componentdiffcross} can be integrated between two scales $Q_{\textrm{min}}$ and $Q_{\textrm{max}}$, which we denote using the notation $\Sigma^{(n_f,\kappa)}$ , which is schematically given as
\begin{equation}
\Sigma^{(n_f,\kappa)}(Q_{\textrm{min}},Q_{\textrm{max}}) = \int_{Q_{\textrm{min}}^2}^{Q_{\textrm{max}}^2}\rm d Q^2\,\frac{\rm d \sigma^{(n_f,\kappa)}}{\rm d Q^2}\,.
\end{equation}
Here we suppress the dependence of all quantities on the renormalization and factorization scales. Expanding the above quantity in terms of $a_s$ we write:
\begin{equation}
\Sigma^{(n_f,\kappa)}_{\textrm{N$^k$LO}}(Q_{\textrm{min}},Q_{\textrm{max}}) = \sum_{l=0}^k a_s^{(n_f)}(\mu_R)^l\,\Sigma^{(n_f,\kappa,l)}(Q_{\textrm{min}},Q_{\textrm{max}})\,.
\end{equation}
After verifying the matching procedure between different flavor schemes, as established in Section \ref{sec:matching_theory}, we assess the effects of power corrections arising from the massive corrections.
\begin{figure}
\centering
\begin{minipage}[t]{1\textwidth}
  \centering  \includegraphics[width=1\linewidth]{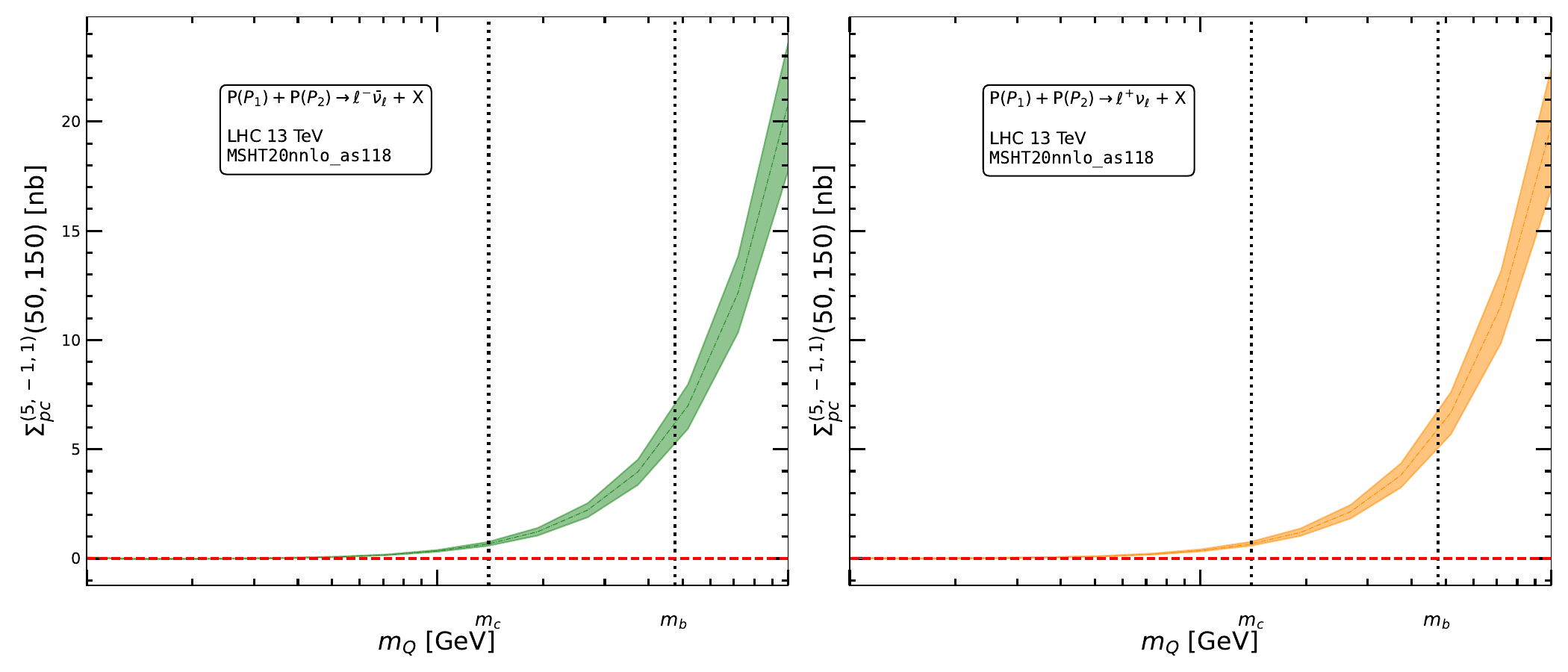}
\end{minipage}
\caption{Power-corrections to the CCDY process up to NLO accuracy for the $Q$-bin (50 GeV, 150 GeV). The bands represent the 7-point variation of $\mu_R$ and $\mu_F$ around the central (dynamic) scale $\mu_R=\mu_F=Q$. We observe that the power corrections vanish in the limit $m_Q \rightarrow 0$.}
\label{fig:CDY_LO_pc}
\end{figure}
For example, in Figures~\ref{fig:CDY_LO_pc} and~\ref{fig:enter-label3}, we present the power corrections for CCDY at NLO the $gg$ and $q\bar{q}$ channels: at NNLO for NCDY. The plots demonstrate that the power correction terms vanish in the limit 
$m_Q \rightarrow 0$, confirming the expected behavior. In addition to validating the matching procedure, we also display the scale variation of the power correction terms separately for the $gg$ and $q\bar{q}$ channels. This is shown within the LHCb fiducial region, for $Q \in [80, 105 ]$ GeV in Figure~\ref{fig:enter-label3}, and for $Q \in [50, 150 ]$ GeV in Figure~\ref{fig:CDY_LO_pc}.



\begin{figure}
    \centering
    \includegraphics[width=1.0\linewidth]{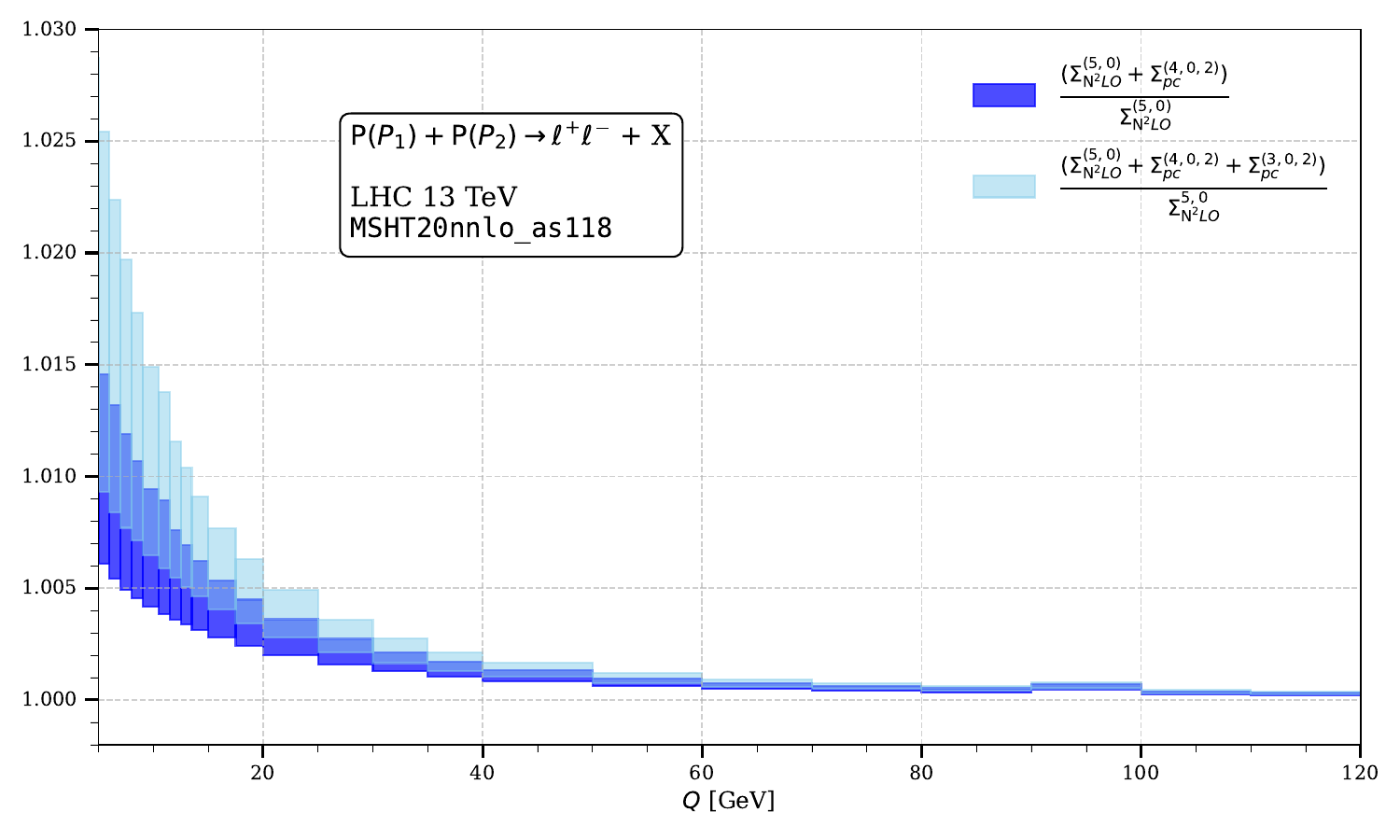}
    \caption{The plot shows the impact of the power corrections on the invariant-mass distribution of the NCDY process. The cross sections are obtained with {\texttt{MSHTnnlo\_as118}} PDF set.}
    \label{fig:sigma_extreme_PC}
\end{figure}

We also plot the impact of the extracted power correction terms over the entire fiducial region $Q \in [4,120]$ GeV in Figure~\ref{fig:sigma_extreme_PC}, comparing them with respect to the NNLO NCDY cross-section.
We observe that the 3FS contributes more than the 4FS. This implies that the combined mass effects of charm and bottom has more effects on the cross section, as compared with only bottom mass effects. This is particularly true for the low-$Q$ region. We also provide a numerical estimate of the power corrections for physical quark mass values, using 
$m_b =4.75$ GeV and $m_c = 1.4$ GeV, as shown in Table~\ref{Table:Fiducial-CDY}.


\begin{table}[!h]
\renewcommand{\arraystretch}{2.2}
\begin{tabular}{|c|c|c|c|}
\hline
Process & $\Sigma^{(n_f,0)}(Q_{min},Q_{max})$              & Prediction (pb)   \\ \hline
 &$\Sigma_{\text{N}^2\text{LO}}^{(5,0)}(80,105)$            & $1824.63$\\ 
NCDY &$\Sigma_{pc}^{(4,0,2)}(80,105)$          & $0.965396$       \\ 
&$\Sigma_{pc}^{(3,0,2)}(80,105)$                        & $0.088011$                \\ \hline
&$\Sigma_{\text{N}\text{LO}}^{(5,+1)}(50,150)$            & $11482.76$\\ 
CCDY &$\Sigma_{pc}^{(3,+1,1)}(50,150)$                        & $0.662841$                \\ 
 \hline
 &$\Sigma_{\text{N}\text{LO}}^{(5,-1)}(50,150)$            & $8524.48$\\ 
CCDY &$\Sigma_{pc}^{(3,-1,1)}(50,150)$                        & $0.662637$                \\ 
 \hline
\end{tabular}
	\caption{Predictions from the 5FS and the massive power-corrections at  are shown for the central scale $\mu_R=\mu_F=Q$ for physical values of quark masses.}
\label{Table:Fiducial-CDY}
\end{table}


\section{Conclusion}\label{sec13}
In this contribution, we analyze the impact of power correction terms arising from bottom and charm quark masses at NNLO for the NCDY process and at NLO for CCDY. In the low-$Q$ region, the corrections for NCDY can be as large as 2.5\%–5\% within the range $Q \in [4,30]$ GeV. For CCDY, the power corrections are much smaller—around 0.006\% in the range $Q \in [50,150]$ GeV. In the process of determining these corrections, we also verify the consistency of the matching scheme presented here.  


\backmatter





\section*{Declarations}
\begin{itemize}
\item Funding
The work of EC is funded by the ERC grant
101043686 ‘LoCoMotive’.
\item Conflict of interest/Competing interests: Not applicable
\item Ethics approval and consent to participate : Views
and opinions expressed are however those of the author(s) only and do not necessarily reflect
those of the European Union or the European Research Council. Neither the European Union
nor the granting authority can be held responsible for them.
\item Consent for publication: The authors provide their consent for publication of the data
\item Data Availability Statement: No Data associated in the manuscript.
\item Materials availability: Not applicable
\item Code availability: Not applicable 
\item Author contribution: The authors contributed equally to this manuscript.
\end{itemize}

\noindent

\begin{appendices}




\end{appendices}


\bibliography{sn-bibliography}

@article{Duhr:2020kzd,
    author = "Duhr, Claude and Dulat, Falko and Hirschi, Valentin and Mistlberger, Bernhard",
    title = "{Higgs production in bottom quark fusion: matching the 4- and 5-flavour schemes to third order in the strong coupling}",
    eprint = "2004.04752",
    archivePrefix = "arXiv",
    primaryClass = "hep-ph",
    reportNumber = "CERN-TH-2020-051, MIT-CTP/5191",
    doi = "10.1007/JHEP08(2020)017",
    journal = "JHEP",
    volume = "08",
    number = "08",
    pages = "017",
    year = "2020"
}

@article{Buza:1996wv,
    author = "Buza, M. and Matiounine, Y. and Smith, J. and van Neerven, W. L.",
    title = "{Charm electroproduction viewed in the variable flavor number scheme versus fixed order perturbation theory}",
    eprint = "hep-ph/9612398",
    archivePrefix = "arXiv",
    reportNumber = "NIKHEF-96-027, ITP-SB-96-66, DESY-96-258, INLO-PUB-22-96",
    doi = "10.1007/BF01245820",
    journal = "Eur. Phys. J. C",
    volume = "1",
    pages = "301--320",
    year = "1998"
}

@article{Gauld:2021zmq,
    author = "Gauld, Rhorry",
    title = "{A massive variable flavour number scheme for the Drell-Yan process}",
    eprint = "2107.01226",
    archivePrefix = "arXiv",
    primaryClass = "hep-ph",
    reportNumber = "NIKHEF 2021-14",
    doi = "10.21468/SciPostPhys.12.1.024",
    journal = "SciPost Phys.",
    volume = "12",
    number = "1",
    pages = "024",
    year = "2022"
}

@article{Behring:2020uzq,
    author = {Behring, Arnd and Bizo\'n, Wojciech and Caola, Fabrizio and Melnikov, Kirill and R\"ontsch, Raoul},
    title = "{Bottom quark mass effects in associated $WH$ production with the $H \to b\bar{b}$ decay through NNLO QCD}",
    eprint = "2003.08321",
    archivePrefix = "arXiv",
    primaryClass = "hep-ph",
    reportNumber = "TTP20-011, P3H-20-009, CERN-TH-2020-043",
    doi = "10.1103/PhysRevD.101.114012",
    journal = "Phys. Rev. D",
    volume = "101",
    number = "11",
    pages = "114012",
    year = "2020"
}

@article{Baglio:2022wzu,
    author = "Baglio, Julien and Duhr, Claude and Mistlberger, Bernhard and Szafron, Robert",
    title = "{Inclusive production cross sections at N$^{3}$LO}",
    eprint = "2209.06138",
    archivePrefix = "arXiv",
    primaryClass = "hep-ph",
    reportNumber = "CERN-TH-2022-109, SLAC-PUB-17699, BONN-TH-2022-22",
    doi = "10.1007/JHEP12(2022)066",
    journal = "JHEP",
    volume = "12",
    pages = "066",
    year = "2022"
}

@article{Chetyrkin:1997sg,
    author = "Chetyrkin, K. G. and Kniehl, Bernd A. and Steinhauser, M.",
    title = "{Strong coupling constant with flavor thresholds at four loops in the MS scheme}",
    eprint = "hep-ph/9706430",
    archivePrefix = "arXiv",
    reportNumber = "MPI-PHT-97-025",
    doi = "10.1103/PhysRevLett.79.2184",
    journal = "Phys. Rev. Lett.",
    volume = "79",
    pages = "2184--2187",
    year = "1997"
}

@article{Duhr:2019tlz,
    author = "Duhr, Claude and Dulat, Falko",
    title = "{PolyLogTools \textemdash{} polylogs for the masses}",
    eprint = "1904.07279",
    archivePrefix = "arXiv",
    primaryClass = "hep-th",
    reportNumber = "CP3-19-17, CERN-TH-2019-045, SLAC-PUB-17423",
    doi = "10.1007/JHEP08(2019)135",
    journal = "JHEP",
    volume = "08",
    pages = "135",
    year = "2019"
}

@article{Duhr:2021vwj,
    author = "Duhr, Claude and Mistlberger, Bernhard",
    title = "{Lepton-pair production at hadron colliders at N$^{3}$LO in QCD}",
    eprint = "2111.10379",
    archivePrefix = "arXiv",
    primaryClass = "hep-ph",
    reportNumber = "BONN-TH-2021-12, SLAC-PUB-17632",
    doi = "10.1007/JHEP03(2022)116",
    journal = "JHEP",
    volume = "03",
    pages = "116",
    year = "2022"
}

@article{Forte:2016sja,
    author = "Forte, Stefano and Napoletano, Davide and Ubiali, Maria",
    title = "{Higgs production in bottom-quark fusion: matching beyond leading order}",
    eprint = "1607.00389",
    archivePrefix = "arXiv",
    primaryClass = "hep-ph",
    reportNumber = "TIF-UNIMI-2016-6, IPPP-16-61, CAVENDISH-HEP-16-11, TIF-UNIMI-2016-6; IPPP/16/61; Cavendish-HEP-16/11",
    doi = "10.1016/j.physletb.2016.10.040",
    journal = "Phys. Lett. B",
    volume = "763",
    pages = "190--196",
    year = "2016"
}

@article{Forte:2018ovl,
    author = "Forte, Stefano and Napoletano, Davide and Ubiali, Maria",
    title = "{$Z$ boson production in bottom-quark fusion: a study of $b$-mass effects beyond leading order}",
    eprint = "1803.10248",
    archivePrefix = "arXiv",
    primaryClass = "hep-ph",
    doi = "10.1140/epjc/s10052-018-6414-8",
    journal = "Eur. Phys. J. C",
    volume = "78",
    number = "11",
    pages = "932",
    year = "2018"
}

@article{Bauer:2000cp,
    author = "Bauer, Christian W. and Frink, Alexander and Kreckel, Richard",
    title = "{Introduction to the GiNaC framework for symbolic computation within the C++ programming language}",
    eprint = "cs/0004015",
    archivePrefix = "arXiv",
    reportNumber = "MZ-TH-00-17",
    doi = "10.1006/jsco.2001.0494",
    journal = "J. Symb. Comput.",
    volume = "33",
    pages = "1--12",
    year = "2002"
}

@article{Campbell:2015qma,
    author = "Campbell, John M. and Ellis, R. Keith and Giele, Walter T.",
    title = "{A Multi-Threaded Version of MCFM}",
    eprint = "1503.06182",
    archivePrefix = "arXiv",
    primaryClass = "physics.comp-ph",
    reportNumber = "FERMILAB-PUB-15-043-T",
    doi = "10.1140/epjc/s10052-015-3461-2",
    journal = "Eur. Phys. J. C",
    volume = "75",
    number = "6",
    pages = "246",
    year = "2015"
}

@article{Campbell:2019dru,
	archiveprefix = {arXiv},
	author = {Campbell, John and Neumann, Tobias},
	doi = {10.1007/JHEP12(2019)034},
	eprint = {1909.09117},
	journal = {JHEP},
	pages = {034},
	primaryclass = {hep-ph},
	reportnumber = {FERMILAB-PUB-19-477-T, IIT-CAPP-19-03},
	title = {{Precision Phenomenology with MCFM}},
	volume = {12},
	year = {2019}}

@article{Cieri:2018sfk,
    author = "Cieri, Leandro and Ferrera, Giancarlo and Sborlini, German F. R.",
    title = "{Combining QED and QCD transverse-momentum resummation for Z boson production at hadron colliders}",
    eprint = "1805.11948",
    archivePrefix = "arXiv",
    primaryClass = "hep-ph",
    reportNumber = "TIF-UNIMI-2018-4",
    doi = "10.1007/JHEP08(2018)165",
    journal = "JHEP",
    volume = "08",
    pages = "165",
    year = "2018"
}

@article{Autieri:2023xme,
    author = "Autieri, Andrea and Cieri, Leandro and Ferrera, Giancarlo and Sborlini, German F. R.",
    title = "{Combining QED and QCD transverse-momentum resummation for W and Z boson production at hadron colliders}",
    eprint = "2302.05403",
    archivePrefix = "arXiv",
    primaryClass = "hep-ph",
    reportNumber = "IFIC/23-06, FTUV-22-1126.2949",
    doi = "10.1007/JHEP07(2023)104",
    journal = "JHEP",
    volume = "07",
    pages = "104",
    year = "2023"
}

@article{Bagnaschi:2018dnh,
    author = "Bagnaschi, Emanuele and Maltoni, Fabio and Vicini, Alessandro and Zaro, Marco",
    title = "{Lepton-pair production in association with a $ b\overline{b} $ pair and the determination of the $W$ boson mass}",
    eprint = "1803.04336",
    archivePrefix = "arXiv",
    primaryClass = "hep-ph",
    reportNumber = "DESY 18-024, CP3-18-16, NIKHEF/2018-008, TIF-UNIMI 2018-2, DESY-18-024, NIKHEF-2018-008, TIF-UNIMI-2018-2",
    doi = "10.1007/JHEP07(2018)101",
    journal = "JHEP",
    volume = "07",
    pages = "101",
    year = "2018"
}

@article{Chaubey:2025lan,
    author = "Chaubey, Ekta and Duhr, Claude and Gauld, Rhorry and Mukherjee, Pooja",
    title = "{A comprehensive analysis of Drell-Yan production uncertainties and mass effects at moderate and low dilepton masses}",
    eprint = "2508.08956",
    archivePrefix = "arXiv",
    primaryClass = "hep-ph",
    reportNumber = "BONN-TH-2025-26, DESY-25-096, MPP-2025-153",
    month = "8",
    year = "2025"
}

\end{document}